\documentclass[aps, pra, twocolumn, reprint, nofootinbib, longbibliography]{revtex4-1}

\usepackage{lmodern}
\usepackage{sistyle}
\usepackage{graphicx}
\usepackage{amsmath}
\usepackage{amssymb}
\usepackage{upgreek}
\usepackage{mathrsfs}
\usepackage{verbatim}   % useful for program listings
\usepackage{subfigure}  % use for side-by-side figures
\usepackage{hyperref}
\usepackage{xcolor}
\newcommand{\um}{\upmu\text{m}}
\newcommand{\uW}{\upmu\text{W}}
\newcommand{\SiN}{Si$_3$N$_4$~}
\newcommand{\TEMOO}{TEM$_{00}$~}
\newcommand{\F}{\mathscr{F}}
\newcommand{\meff}{m_\text{eff}}

\begin{document}

\title
  {Ultralow-Noise SiN Trampoline Resonators for Sensing and Optomechanics}

\author{Christoph Reinhardt, Tina M\"{u}ller, Alexandre Bourassa, Jack C. Sankey}
\affiliation{Department of Physics, McGill University, Montr\'{e}al, 
	Qu\'{e}bec, H3A 2T8, Canada}

\email{jack.sankey@mcgill.ca}


\begin{abstract}
	In force sensing, optomechanics, and quantum motion experiments, it is typically advantageous to create lightweight, compliant mechanical elements with the lowest possible force noise. Here we report wafer-scale batch fabrication and characterization of high-aspect-ratio, nanogram-scale \SiN~``trampolines'' having quality factors above $4 \times 10^7$ and ringdown times exceeding five minutes ($1$ mHz linewidth). We measure a thermally limited force noise sensitivity of 16.2$\pm$0.8 aN/Hz$^{1/2}$ at room temperature, with a spring constant ($\sim$1 N/m) 2-5 orders of magnitude larger than those of competing technologies. We also characterize the suitability of these devices for high-finesse cavity readout and optomechanics applications, finding no evidence of surface or bulk optical losses from the processed nitride in a cavity achieving finesse 40,000. These parameters provide access to a single-photon cooperativity $C_0 \sim 8$ in the resolved-sideband limit, wherein a variety of outstanding optomechanics goals become feasible.
\end{abstract}

% possibly not necessary for ACS mode
\maketitle

Advances in nanofabrication over the past decades have enabled the growth and patterning of pristine materials, and the creation of mechanical sensors of extraordinary quality \cite{Poot2012Mechanical}. Cantilevers sensitive to attonewton forces at room temperature have been fabricated from silicon (e.g.~$50$ aN/Hz$^{1/2}$ \cite{Yasumura2000Quality}) and diamond ($26$ aN/Hz$^{1/2}$ \cite{Tao2014Single}) using ``top-down'' techniques, while ``bottom-up'' fabricated devices can in principle achieve below 10 aN/Hz$^{1/2}$ at room temperature (e.g.~approaching $\sim$5~aN/Hz$^{1/2}$ for silicon nanowires \cite{Nichol2008Displacement} or carbon nanotubes \cite{Jensen2008An}), and $1$~zN/Hz$^{1/2}$ at low temperatures (nanotubes \cite{Moser2014Nanotube}). These complementary approaches carry with them an important trade-off: on one hand, bottom-up techniques can assemble fewer atoms into smaller, more sensitive structures, but the technology is comparatively young, and it is more difficult to incorporate additional structures and/or probes. These low-mass objects also tend to have very low spring constants (i.e.~below $\sim$10 mN/m), making them highly susceptible to van der Waals ``sticking'' forces at short distances. On the other hand, top-down devices are currently not as sensitive at low temperature (e.g.~$\sim 500$ zN/Hz$^{1/2}$  for diamond at 93 mK \cite{Tao2014Single}), but are reliably fabricated, are compatible with a wide variety of probes, and naturally integrate with other on-chip systems. Some of their remarkable achievements to date include detection of a single electron spin \cite{Rugar2004Single}, nanoscale clusters of nuclei \cite{Degen2009Nanoscale}, persistent currents in normal metal rings \cite{Castellanos2013Measurement}, and the force noise associated with the quantized nature of light \cite{Purdy2013Observation}. Further, integrating with quantum electronics and / or optical resonators has provided (among other things) access to a regime in which quantum effects play a central role in the mechanical element's motion \cite{OConnell2010Quantum, Teufel2011Sideband, Chan2011Laser, Safavi2012Observation, Purdy2015Optomechanical, Underwood2015Measurement, Meenehan2015Pulsed}. 

The intrinsic force noise of a mechanical system is ultimately determined by its dissipative coupling to the environment \cite{Saulson1990Thermal}. In equilibrium, assuming the average energy flow to and from the environment balances such that the equipartition theorem is satisfied, the force noise density experienced by the mechanical system is
\begin{equation}
	S_{F}= \sqrt{8 \meff k_B T /\tau_m},
\end{equation}
where $T$ is the temperature of the environment, $\meff$ is the participating (effective) mass of the resonator, $\tau_m$ is its amplitude ringdown time, and $k_B$ is the Boltzmann constant. Written this way, it is immediately evident that the fundamental thermal noise floor of a mechanical sensor benefits from a small mass and a long ringdown time.

Here we report wafer-scale batch fabrication of high-aspect-ratio, nanogram-scale \SiN ``trampoline'' resonators with ringdown times $\tau_m$ approaching 6 minutes (1 mHz linewidth) at room temperature. This class of devices, together with Ref.~\cite{Norte2016Mechanical} (submitted simultaneously), consistently achieve an intrinsic force noise below $20$~aN/Hz$^{1/2}$ at room temperature (293 K), and our measured value $S_F$=16.2$\pm$0.8 aN/Hz$^{1/2}$ is similar to what is in principle possible using a single layer of graphene \cite{Kumar2015Ultrasensitive}. Furthermore, this low noise is accompanied by spring constants $K_\text{eff} \sim 1$~N/m that are $\sim$$2$-$5$ orders of magnitude higher than those of competing devices \cite{Poot2012Mechanical, Yasumura2000Quality, Nichol2008Displacement, Tao2014Single, Jensen2008An, Kumar2015Ultrasensitive}, and the $\sim$100$\times$100 $\um^2$ surface area is compatible with the incorporation of additional structures \cite{Norte2016Mechanical}. We also demonstrate their suitability for sensitive interferometric readout and optomechanics applications by positioning an extended membrane (fabricated by the same means) within an optical cavity of finesse $\mathscr{F}=20,000$, finding no evidence of additional bulk or surface optical losses from the processed nitride at telecom wavelengths (1550 nm), consistent with literature \cite{Wilson2009Cavity, Sankey2010Strong}. In fact, for certain membrane positions, the cavity finesse is increased to $40,000$, as expected for a lossless dielectric slab in a single-port cavity. Finally, to set an approximate upper bound on the size of the cavity field required for high-finesse applications, we position a trampoline in a cavity field wide enough that $0.045\%$ of the light falls outside the structure. Consistent with recent simulations \cite{Chang2012Ultrahigh}, we find that the majority of this ``clipped'' light is in many cases recovered by the cavity.

\section{Mechanical Properties}\label{sec:fab}

\begin{figure}[!ht]
	\centering
	\includegraphics[width=0.9\columnwidth]{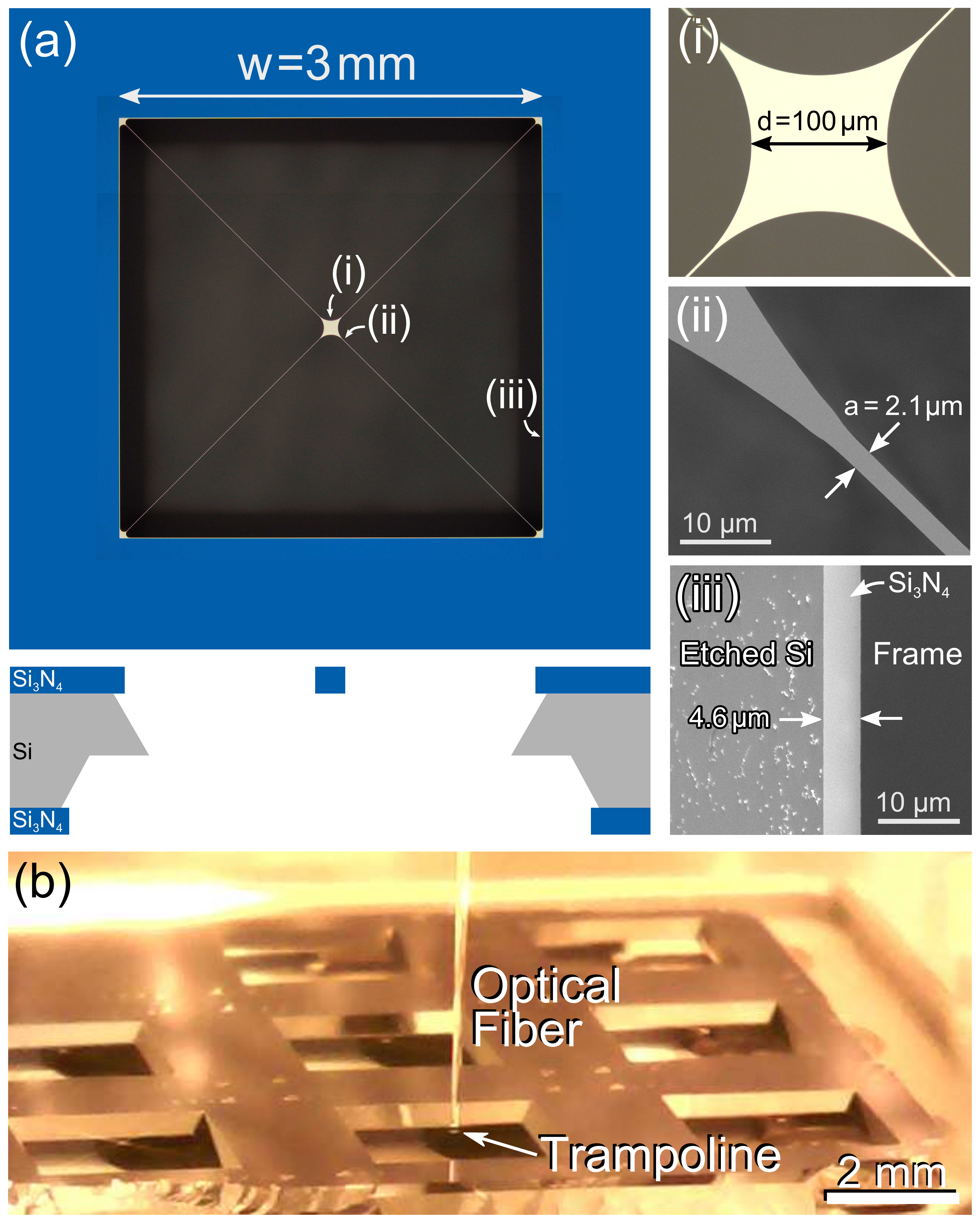}
	\caption{Fabricated \SiN ``trampoline'' resonators. (a) Optical image of the released structure with a window size of $w=3$~mm (upper) and a schematic of its KOH-etched cross-section (lower). Right-hand images show (i) an optical image of the $d=100~\um$-wide central pad, (ii) a scanning electron microscope (SEM) image of the $a=2.1$-$\um$-wide tether (near the pad), and (iii) an SEM image of the 4.6-$\um$-wide overhanging nitride. Left from the overhang is the angled, KOH-etched silicon substrate showing typical roughness and residues. (b) Optical image of devices inside the ultrahigh vacuum (UHV) fiber interferometer.}
	\label{fig1}
\end{figure}

Drawing inspiration from similar structures having embedded Bragg mirrors \cite{Groblacher2009Demonstration, Kleckner2011Optomechanical} and high-$Q_m$ nitride strings \cite{Verbridge2006High}, we pattern single-layer resonators suitable for a ``membrane-in-the-middle'' \cite{Thompson2008Strong} optomechanical geometry. Figure \ref{fig1}(a) shows a typical structure, comprising (i) an 80-nm-thick, 100-$\um$-wide central pad suspended by (ii) 2.1-$\um$-wide tethers. These devices are suspended from a 675-$\um$-thick, (single-side-polished) silicon wafer, upon which 100 nm of stoichiometric \SiN was commercially deposited via low-pressure chemical vapor deposition.\footnote{Note \SiN-coated wafers purchased from University Wafer and Addison Engineering produce similar results.} Nitride on silicon appears blue, and suspended nitride appears yellow. The filleted shapes \cite{Verbridge2006High} of the central nitride pad and corner clamping points ensure that all suspended structures are held flat by the nitride's internal stress (nominally $\sim 1$ GPa), and that regions of concentrated strain in the structure's normal modes are minimized. The fillets are nominally circular; on the central pad their radius defines the pad diameter $d$ and the corner fillets are defined to have a quarter of this radius, to reduce their relative mass. The tethers are long ($2$ mm) to simultaneously increase the mechanical quality factor $Q_m$ \cite{Schmid2011Damping} and decrease the mechanical frequency $\omega_m$, thereby maximizing $\tau_m$ without contributing too much mass. The cross section of the wafer (lower image of Fig.~\ref{fig1}(a), also faintly visible from above) results from the minimum anisotropic KOH etch required to cut a clear-shot window through the silicon. This choice minimizes the region of overhanging nitride (iii), a known source of mechanical dissipation \cite{Ni2012Enhancement, Schmid2011Damping}. The angle of the undercut silicon associated with this etch technique also serves to further increase the rigidity of the supporting frame at the clamping points. Additional fabrication details can be found in Appendix I.

We characterize the structure's mechanical resonances using a fiber interferometer at a vacuum below $10^{-6}$ torr (see Fig.~\ref{fig1}(b)). Laser light is directed along a fiber toward a cleaved tip (positioned within $\sim$100 $\um$ of the trampoline), and the interference between reflections from the cleave and trampoline records the instantaneous displacement. A piezo actuator attached to the stage is used to exert an oscillatory mechanical drive.

\begin{figure}[!ht]
	\centering
	\includegraphics[width=0.95\columnwidth]{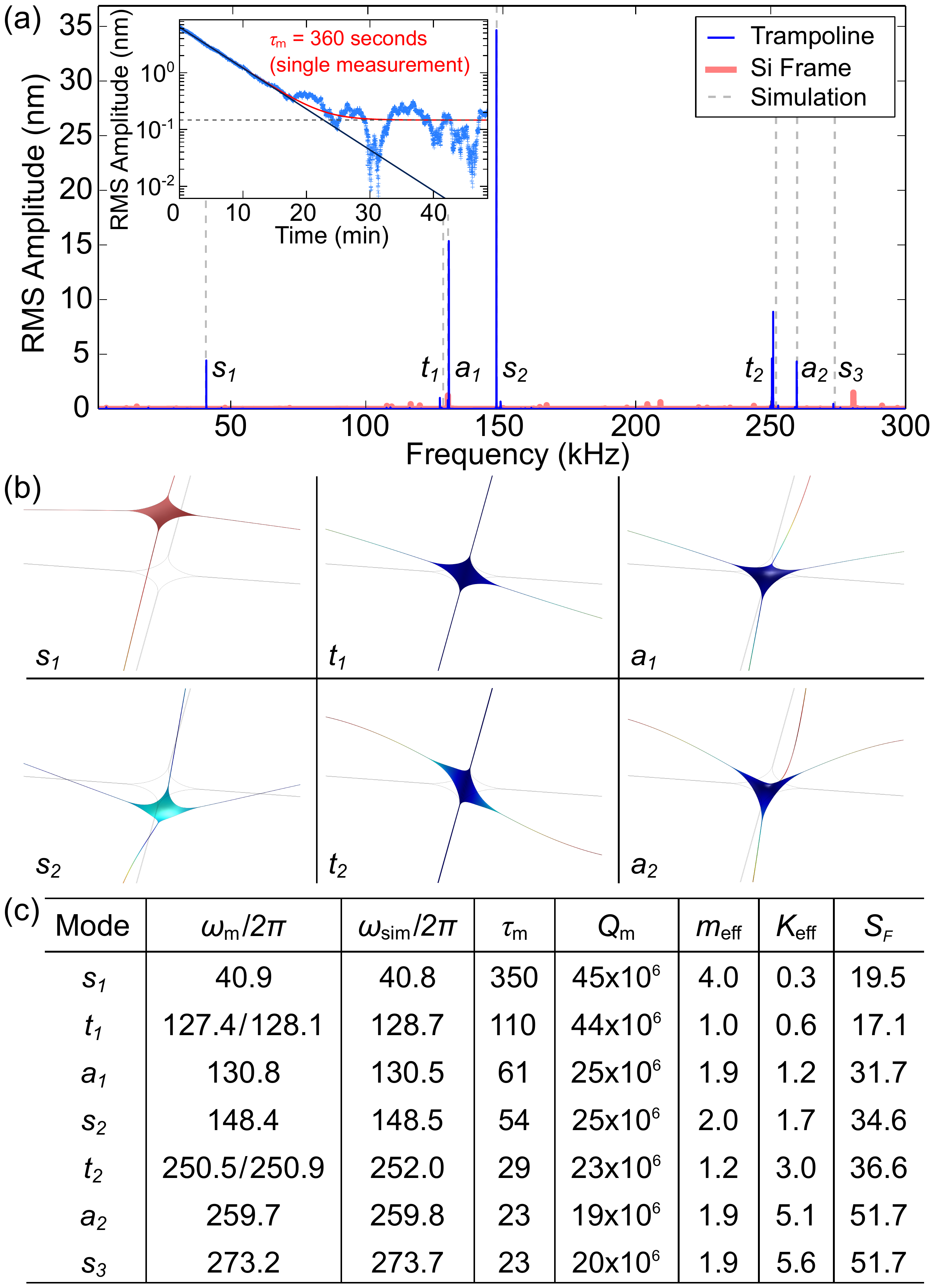}
	\caption{Mechanical modes of a trampoline having lateral dimensions of Fig.~\ref{fig1} and thickness 80 nm, measured with a fiber interferometer operating at wavelength 1550 nm and power 220 $\uW$. (a) Approximate response to piezo drive, showing first nine resonances (thin blue line). Pink line shows the response of the Si frame. Simulated resonance frequencies (dashed gray lines) agree to within $1\%$ of measured values with \SiN parameters density 2700 kg/m$^3$, Young's modulus 250 GPa, Poisson ratio 0.23, and internal stress $0.95$ GPa. Inset shows a ``typical'' ringdown for the fundamental (``symmetric'' $s_1$) mode with fit (red curve) having functional form $\sqrt{(x_0 e^{-t/\tau_m})^2+x_1^2}$, where $x_0$, $\tau_m$, and $x_1$ are allowed to float. Black line shows the ringdown extrapolated from the early data, and gray dashed line shows $x_1$ (run-to-run variation by a factor of $\sim 2$). The ringdown time $\tau_m=350\pm 15$ s (error represents statistical fluctuations of multiple measurements) corresponds to a room temperature force noise $S_{F}=19.5\pm0.5$ aN/Hz$^{1/2}$. (b) Simulated displacement profiles for the ``symmetric'' ($s_i$), ``torsional'' ($t_i$) and ``antisymmetric'' ($a_i$) modes labeled in (a). (c) Measured frequency $\omega_m/2\pi$ (kHz), simulated frequency $\omega_\text{sim}/2\pi$ (kHz), ringdown time $\tau_m$ (s), quality factor $Q_m$, mass $\meff$ (ng), spring constant $K_\text{eff}$ (N/m) and force noise $S_{F}$ (aN/Hz$^{1/2}$) for the first 9 modes. The mass has a $\sim$10\% systematic error due to uncertainty in the thickness and density of the nitride.}
	\label{fig2}
\end{figure}

Figure~\ref{fig2}(a) shows the amplitude of driven oscillations as a function of frequency for the fiber positioned over the nitride pad (blue) and silicon frame (pink). Both curves contain many peaks, and several very strong resonances (labeled) emerge whenever the tip is positioned over the pad. There are a few ways to convincingly identify these as trampoline modes, aside from noting their large response. First, they uniformly exhibit significantly larger quality factors $Q_m > 10^7$ (measured by ringdown; see below), whereas supporting frame resonances exhibit low-amplitude peaks of $Q_m<10^5$. Second, we compare the observed frequencies with those predicted by a finite-element simulation (COMSOL) of our geometry. We simulate the volume of the released nitride in the thin membrane limit, and apply perfectly-clamped boundary conditions along the outer edges of the overhanging nitride. The nitride itself is modeled using the material parameters listed in the caption, and we set its internal stress to \SI{0.95}{GPa}. The resulting normal mode frequencies are indicated with dashed lines in Fig.~\ref{fig2}(a), and the corresponding mode shapes are illustrated in Fig.~\ref{fig2}(b). These parameters reproduce all 9 frequencies of the high-$Q_m$ resonances (i.e. the 7 labeled in Fig.~\ref{fig2}(a), with twofold degeneracies for the ``torsional'' modes $t_1$ and $t_2$) to within 1\% of the observed values. It is worth noting that some peaks in Fig.~\ref{fig1}(a) appear artificially small because we did not let the drive dwell on resonance long enough for the mode to ring up; this requires $>$10 minutes per point, and small temperature drifts shift the resonance by more than the ($\sim$mHz) linewidth during this time.

To determine $Q_m$, we instead perform a mechanical ringdown by suddenly switching off a near-resonant drive and monitoring the amplitude decay. A ``typical'' ringdown is shown in Fig.~\ref{fig2}(a, inset), along with an exponential fit (red) for the 40.9-kHz fundamental (``symmetric'' $s_1$) mode. Due to the thermally-driven noise of the mode (visible at the end of the ringdown and discussed below) repeated fit values span $\tau_m = 350 \pm 15$ seconds, corresponding to $Q_m=45\pm2 \times 10^6$ and a room-temperature thermal force noise of $19.5\pm0.5$ aN/Hz$^{1/2}$. 

The fit values for the higher-order mechanical modes are listed in Fig.~\ref{fig2}(c). Of note, the first ``torsional'' mode $t_1$ achieves a marginally lower force noise, and may in fact be more useful for some of the classical sensing geometries suggested in Section \ref{sec:discussion}. For reference, table \ref{table1} also lists the properties of other trampolines having similar pad diameters $d$ and tether widths $a$ but different window sizes $w$ (see Ref.~\cite{Norte2016Mechanical} for additional parameter variations). In agreement with previous studies, we find larger $Q_m$ for longer tethers \cite{Schmid2011Damping, Verbridge2006High}. 

\begin{table}
	\begin{center}
		\begin{tabular}{ c | c | c | c | c | c | c | c | c}
			\hline
			$w$ & $d$ & $a$ & $\omega_m/2\pi$ & $\tau_m$ & $Q_m$ & $\meff$ & $K_\text{eff}$ & $S_{F}$ \\ \hline
			375  & 100 & 1.4 & 196.3 & 8 & 5$\times10^6$  & 2.5 & 3.8 & 101.8 \\ 
			750  & 100 & 1.6 & 101.9 & 25  & 8$\times10^6$  & 3.0 & 1.2 & 62.3  \\
			2400 & 90  & 2.0 & 51.5  & 238 & 39$\times10^6$ & 3.7 & 0.4 & 22.4  \\
			3000 & 100 & 2.1 & 40.9  & 350 & 45$\times10^6$ & 4.0 & 0.3 & 19.5  \\
			\hline		
		\end{tabular}
	\end{center}
	\caption{Frequency $\omega_m/2\pi$ (kHz), ringdown time $\tau_m$ (s), quality factor $Q_m$ effective mass $\meff$ (ng), spring constant $K_\text{eff}$ (N/m), and force noise $S_{F}$ (aN/Hz$^{1/2}$) for the fundamental ($s_1$) mechanical resonance of trampolines having varied window size $w$ ($\um$), pad diameter $d$ ($\um$), and tether width $a$ ($\um$). }
	\label{table1}
\end{table}

\begin{figure}[!ht]
	\centering
	\includegraphics[width=0.95\columnwidth]{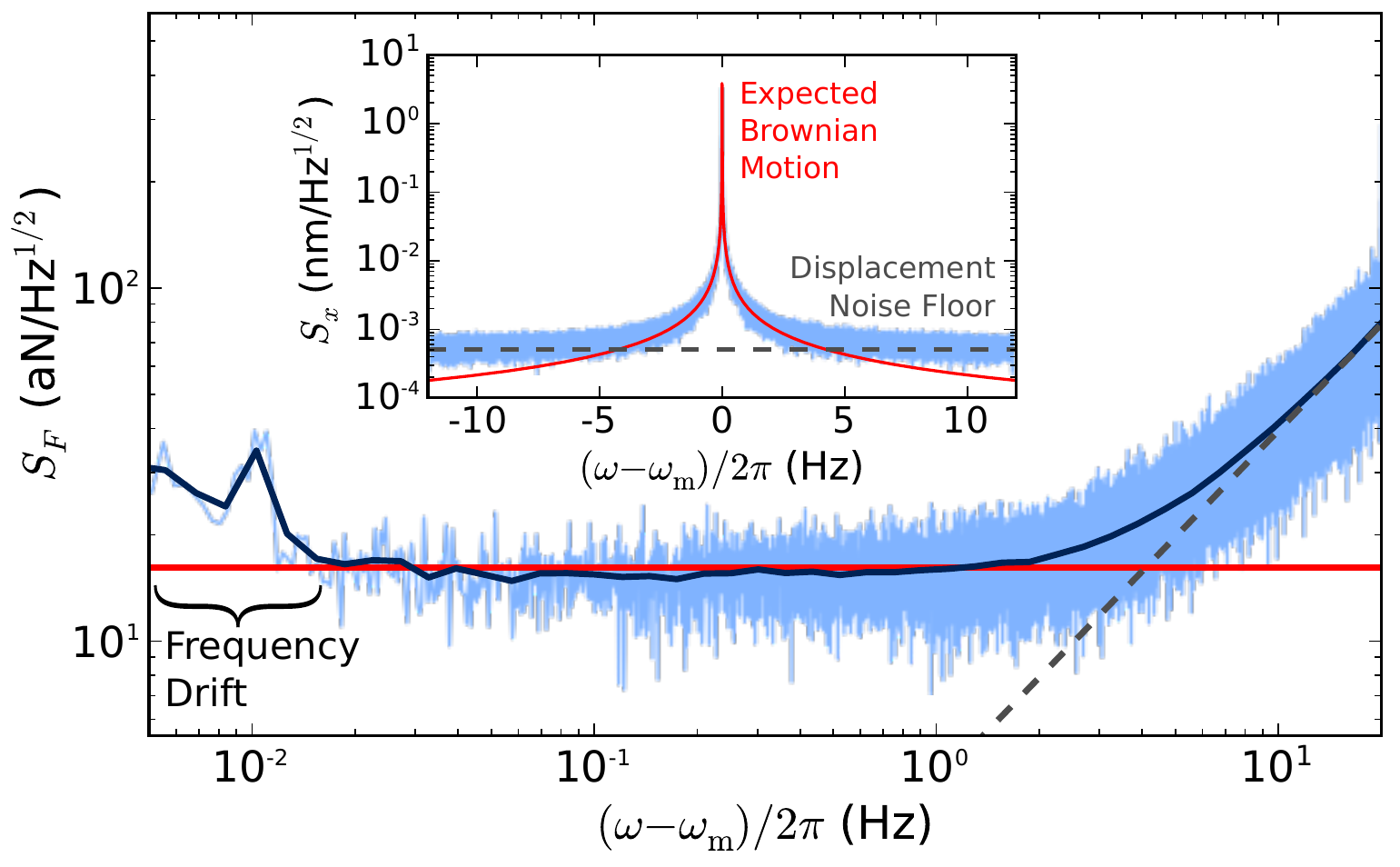}
	\caption{Force noise measurement for the $s_1$ mode of a 44 nm-thick trampoline. Inset shows the displacement noise spectrum $S_{x,\text{obs}}$ (blue) observed for a laser power of 50 $\uW$, along with the spectrum expected for the measured device parameters $S_x$ (red) and the displacement noise floor of the interferometer (dashed, $509 \pm 44$ pm/Hz$^{1/2}$). Main panel shows the force noise spectrum $S_{F,\text{obs}} = S_{x,\text{obs}} S_F/S_x$ (blue), consistent (within a $\sim 5\%$ systematic calibration error) with the expected $16.2 \pm 0.8$ aN/Hz$^{1/2}$ (red) over many thousands of linewidths. Dark blue line is the same data ``coarsened'' by averaging together points within 10\% of each other. Dashed line again indicates the displacement noise floor.} 
	\label{fig3}
\end{figure}

Devices of this scale typically have little trouble achieving the inferred room-temperature force sensitivities. To verify this is, we measure the force noise spectrum of a similar device (Fig.~\ref{fig3}), having the same lateral dimensions as that of Figs.~\ref{fig1} and \ref{fig2}, but with a thickness of 44 nm, owing to a more aggressive HF etch. The fundamental mode frequency $\omega_m = 2\pi \times$ 41.4 kHz, mass $\meff$ = 2.3 ng, and ringdown time $\tau_m=285$ seconds of this device correspond to a thermal force noise of $S_{F}=16.2$ aN/Hz$^{1/2}$ and root-mean-squared displacement $x_\text{RMS}=161$ pm. Letting the system evolve without drive for 38 hours, we observe $x_\text{RMS,\text{obs}}=165 \pm 5$ pm, which agrees with the expected value. Figure \ref{fig3}(inset) shows the displacement noise spectrum $S_{x,\text{obs}}$ for this data (blue data), which follows the expected form \cite{Saulson1990Thermal},
\begin{equation}
S_x = \frac{2 \tau_m k_B T}{\meff \omega_m^2 \left[1+(\omega-\omega_m)^2 \tau_m^2\right]}
\end{equation} 
(red curve, not a fit), before the displacement noise floor dominates above $\sim 4$ Hz from resonance. Since the displacement noise spectrum $S_x$ is just the (white) force noise spectrum $S_F$ ``filtered'' by the harmonic oscillator susceptibility, we can extract the force noise spectrum $S_{F,\text{obs}}$ by multiplying $S_{x,\text{obs}}$ by the ratio $S_F/S_x$, the result of which is plotted in the main panel. Near resonance (within 20 mHz), the noise is limited by temperature-induced drift in the mechanical frequency, and above 20 mHz, we observe a noise floor consistent with $S_F$ over many thousands of mechanical linewidths. This illustrates that these trampolines should present no surprising technical challenges in achieving the inferred sensitivities.\footnote{Note our fiber interferometer was constructed without any consideration to thermal stability or vibration isolation: devices rest on a piezo stage fixed to a stainless plate; this rests directly on a vacuum flange, and the whole chamber is supported by metal blocks on a work bench.}

\section{Optical Properties}\label{sec:optical}

\begin{figure}[!ht]
	\centering
	\includegraphics[width=0.95\columnwidth]{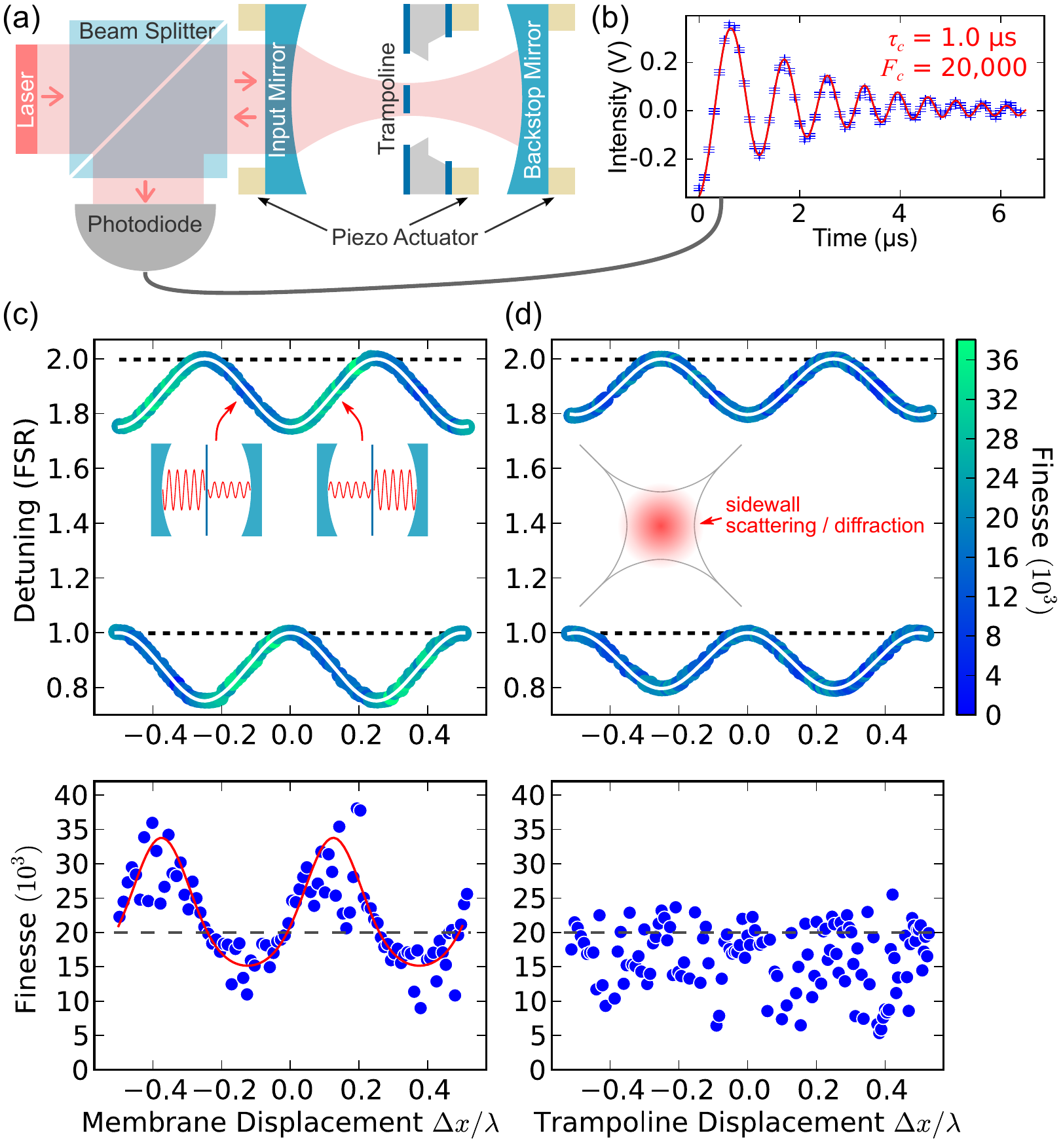}
	\caption{Optical properties of fabricated nitride. (a) Schematic. An extended membrane or trampoline resonator is positioned near the waist of a Fabry-P\'{e}rot cavity. Input mirror reflectivity $|r_i|^2 \approx 0.9997$ and ``backstop'' reflectivity $|r_b|^2 > 0.999993$. (b) Length-swept (empty) cavity optical ringdown: after passing through resonance, light exiting the cavity beats with the prompt reflection, resulting in power fluctuations $a e^{-t/2\tau_c}\cos\left[\left(\omega_0+ b t\right)t+c\right]$ for fit (red curve) parameters $a$, $\omega_0$, $b$, $c$, and power ringdown time $\tau_c=1.00\pm0.05~\upmu$s  (finesse $\F=20,100\pm1,000$, error represents statistical fluctuations of multiple measurements). (c, upper) Finesse and cavity mode detuning versus the displacement of an extended membrane. White curves show fit (see text), and solid dashed lines show the approximate empty cavity resonance frequencies. Inset shows qualitative sketch of left and right cavity modes (c, lower) Comparison of finesse (from the topmost resonance) with prediction for a lossless membrane (red). (d) Same as (c) but for a patterned trampoline of width $d=200~\um$, in this case, the trampoline's effective reflectivity is $|r_t|=0.31\pm0.01$. Inset shows a qualitative sketch of the cavity cross-section at the trampoline.}
	\label{fig4}
	
\end{figure}

While the high performance of these trampolines makes them excellent candidates for mechanical sensing and dissipation studies, we also wish to use them for precision interferometry and optomechanics experiments. To this end we characterize their optical performance by measuring their effect on a high-finesse cavity. Figure \ref{fig4}(a) shows a schematic of the test setup: two high-reflectivity mirrors (2.5 cm radius of curvature) form a Fabry-P\'{e}rot cavity of length $L=4.7$ cm, which, at our operating wavelength $\lambda=1550$ nm, yields a \TEMOO optical mode full-waist $2\sigma=110~\um$ and a free spectral range (FSR) of 3.2~GHz. The input mirror is designed to have a ``modest'' reflectivity of $\approx 0.9997$ while the ``backstop'' (right-hand) mirror reflectivity exceeds $0.999993$, forcing the majority of cavity light to exit through the input mirror. The resulting bare cavity finesse $\F=20,100 \pm 1,000$, as measured by a swept-cavity ringdown \cite{He2001Optical} (see Fig.~\ref{fig4}(b)). 

Unpatterned \SiN membranes fabricated elsewhere have been shown to exhibit very little optical loss \cite{Wilson2009Cavity,Sankey2010Strong}, and in particular, the bound placed on the imaginary index $\text{Im} [n] < 1.5 \times 10^{-6}$ \cite{Sankey2010Strong} would in principle make these structures compatible with a cavity finesse $\mathscr{F}>10^6$, even when positioned at an antinode of the intracavity field. To test whether our fabrication protocol introduces additional bulk absorption or surface losses, we first align a similarly fabricated, extended membrane near the waist of the cavity. The cavity length $L$ is then rapidly swept (symmetrically about the membrane) while the membrane's displacement $\Delta x$ is slowly stepped, producing ringdown curves like Fig.~\ref{fig4}(b) whenever the cavity passes through resonance. We also record the piezo voltages at which ringdown events occur, creating a map of resonant lengths and membrane displacements. Piezo nonlinearity and creep, combined with temperature drift, results in a smoothly distorted and slightly sheered map. To eliminate this, we simultaneously fit the resonant values to their known (periodic) functional dependence on $\Delta x$ and $L$ \cite{Jayich2008Dispersive}, incorporating fourth-order polynomial distortion and linear sheer correction terms. Doing so allows us to extract the cavity detunings induced by the membrane, along with the membrane's reflectivity $|r_m| = 0.38 \pm 0.01$. Using a lower-order polynomial does not significantly change our result, but minor systematics do eventually become visible. Note this value of $|r_m|$ corresponds to that expected for a \SiN (refractive index 2.0) slab of thickness $72 \pm 2$ nm, which is smaller than the nominal value of 80 nm. However, this scheme is known for its systematic underestimate of $|r_m|$ \cite{Sankey2010Strong}, which is attributed to slight misalignment of the membrane and / or level repulsion between the \TEMOO and higher-order transverse modes of the cavity, both of which tend to flatten the sinusoidal perturbation. 

The finesse (color scale and lower plot of Fig.~\ref{fig4}(c)) is found to oscillate with position, in fact achieving a higher value than is possible with the cavity mirrors alone. This can be readily understood by viewing the membrane as ``one more dielectric layer'' of the input mirror that, with the proper air-gap, enhances its reflectivity. A transfer matrix theory \cite{Jayich2008Dispersive} assuming zero optical loss in the membrane (red curve, zero free parameters) reproduces the oscillations. This implies that, so long as the optical mode waist $2\sigma$ is sufficiently small compared to the diameter $d$ of the pad, it should readily achieve a cavity finesse of 40,000 or higher. Note, as observed previously \cite{Sankey2010Strong}, the error bar on individual finesse measurements is significantly smaller than the fluctuations in Fig.~\ref{fig4}; the larger, non-statistical variations are known to arise from membrane-mediated hybridization between the \TEMOO mode and higher-order modes of the cavity (each having its own value of finesse) whenever they approach degeneracy.

Finally, in an effort to place an approximate upper bound on the cavity mode diameter required to achieve this finesse with a patterned device, we replace the membrane with trampoline having a pad diameter $d=200~\um$, such that $\sim 0.045 \%$ of the cavity light (mode diameter $2\sigma=110~\um$) does not land on the structure. If we naively assume this light is lost from the cavity, the finesse would be limited to $7,000$. However, simulations of a similar geometry \cite{Chang2012Ultrahigh} suggest higher value, since the end mirrors can collect and recycle some of the scattered light. As shown in Fig.~\ref{fig4}(d), despite these ``clipping'' losses, a finesse equal to the empty-cavity value of 20,000 is achievable within a short distance of any trampoline position, even near the antinodes of the intracavity field. Clipping effects are still evident, however: the regions of boosted finesse have vanished, and the rapid finesse variations dip to much lower values. This is consistent with an intuition that sidewall scattering further breaks the symmetry of the cavity, increasing the \TEMOO mode's coupling to even higher-order, lossier transverse modes.

\section{Discussion}\label{sec:discussion}

We have fabricated, using scalable top-down techniques, sensitive mechanical systems that are compatible with high-finesse optics. Their low dissipation rates make them excellent candidates for studies of dissipation mechanisms \cite{Chakram2014Dissipation, Villanueva2014Evidence}, and their high stiffness and low force noise makes them well suited for classical sensing applications. It is hard to predict what form the latter might take, but in the simplest case we envision capacitive \cite{Andrews2014Bidirectional, Yuan2015Silicon} or fiber \cite{Rasool2010A} readout from within the silicon etch pit, and a sharp tip (or other probe) fabricated upon the top surface. Alternatively, if high-finesse readout is required (this would boost the thermally-limited bandwidth of Fig.~\ref{fig3}), one could position a probe at the edge of the central pad or upon the tethers, far from any light fields, employ a fiber cavity \cite{Flowers2012Fiber}, and / or exploit the $t_1$ mode, which has the same force sensitivity but a larger tether displacement and spring constant. Furthermore, if $Q_m$ follows the trend for nitride, namely increasing by a factor of 10-100 at low temperature \cite{Zwickl2008High,Yuan2015Silicon}, these devices could in principle achieve $\sim 14$ zN/Hz$^{1/2}$ at 14 mK \cite{Yuan2015Silicon}, a value approaching that of a nanotube \cite{Moser2013Ultrasensitive, Moser2014Nanotube}, but with a significantly larger, stiffer platform amenable to the incorporation of additional circuitry and probes.

The compatibility with high-finesse optics, together with the long ringdown time of mode $s_1$, also provides access to parameter regimes of central interest in the field of optomechanics. One figure of merit is the single-photon cooperativity $C_0$ \cite{Aspelmeyer2014Cavity}, which can be written
\begin{equation}
	C_0=\frac{4\pi\hbar c r_m^2\tau_m \mathscr{F}}{\lambda^2 L\meff\omega_m},
\end{equation}
for this geometry. This unitless parameter provides a measure of how strongly cavity light at the single-photon level can affect the mechanical system. For example, when $C_0$ reaches unity, a cooling laser (in the resolved-sideband limit) having an average intensity of a single cavity photon will provide a dissipation rate equal to that of the bare mechanical element. Interestingly, the demonstrated cavity parameters ($L=4.7$ cm, $\mathscr{F}=40,000$, $\lambda=1550$ nm) and trampoline parameters ($\meff=4.0$~ng, $\tau_m=6.0$~min, $\omega_m=2\pi \times 40.9$ kHz, $r_m$=0.4) correspond to a single-photon cooperativity $C_0 \sim 8$ in the resolved-sideband limit (``resolved'' in the sense that the back-action-limited cooling would result in an average phonon occupancy $\bar{n}_m=0.2<1$ \cite{Wilson2007Theory, Marquardt2007Quantum}). At this value of $C_0$, extraordinarily low levels of light will profoundly influence the trampoline's motion. Equally interestingly, if the trampoline is simultaneously laser cooled to the back-action limit \cite{Peterson2016Laser} and mechanically driven to an amplitude of $\sim$5 nm (i.e. as in Fig.~\ref{fig2}(a)), even the gentle quadratic optomechanical coupling found at a node or antinode \cite{Thompson2008Strong} would be sufficient to perform a quantum nondemolition (QND) readout of the trampoline's phonon shot noise \cite{Clerk2010Quantum} with a signal-to-noise ratio of $\sim 170$. Importantly, such a scheme is inherently compatible with a single-port optical cavity such as the one employed in Fig.~\ref{fig4}, as required by the theory  \cite{Clerk2010Quantum}. This avoids the need to find clever ways to catch and recycle cavity light from one of the two ports found in other systems such as avoided crossings \cite{Sankey2010Strong}, wherein the requirements are significantly more stringent \cite{Miao2009Standard}. Finally, though these devices are not optimized to benefit from the $Q_m$-enhancement of partial levitation \cite{Ni2012Enhancement,Chang2012Ultrahigh,Muller2015Enhanced}, a finite-element simulation (as in, e.g., Refs.~\cite{Chang2012Ultrahigh, Muller2015Enhanced}) predicts that $Q_m$ can be boosted by a factor of $\sim 2.5$ when trapped to $\omega_m \sim 2\pi\times 100$ kHz, thereby achieving $Q_m>10^8$. In this case, there would be an average of $\bar{n}_m = k_B T/\hbar \omega_m \sim 6 \times 10^7$ thermal phonons in the mode at room temperature. This meets the requirement $\bar{n}_m < Q_m$ for laser cooling to the quantum mechanical ground state \cite{Marquardt2007Quantum, Wilson2007Theory}.

As mentioned above, another group \cite{Norte2016Mechanical} simultaneously reported trampoline structures of similarly high mechanical performance, finding that thinning these devices tends to further increase $Q_m$, with one 20-nm-thick device exhibiting $Q_m=9.8\times 10^7$. They furthermore demonstrate the incorporation of a photonic crystal reflector, finding that this addition generally does not affect $Q_m$. These two results agree with and complement one another. Trampolines much thinner than $\sim$100 nm \emph{without} photonic crystals suffer from reduced reflectivity that, despite the correspondingly smaller mass, reduces the overall optomechanical coupling rate \cite{AspelmeyerEDITOR2014Cavity, Aspelmeyer2014Cavity}. For example, despite its smaller mass, the correspondingly lower reflectivity $r_m$ of the 44-nm-thick device (Fig.~\ref{fig3}) results in a cooperativity of just $C_0 \sim 4$. For the high-finesse cavity demonstrated in Fig.~\ref{fig4}, the $\sim$80-nm-thick nitride corresponds to $r_m$ within $\sim 30\%$ of the maximum for a dielectric slab, thereby achieving a \emph{linear} optomechanical coupling within a factor of $\sim$2 of the maximum possible. On the other hand, a photonic crystal reflector can significantly boost the \emph{quadratic} coupling \cite{Stambaugh2015From} while still in principle maintaining a single-port cavity. This provides a promising route toward resolving individual quantum jumps between the phonon number states of the trampoline \cite{Thompson2008Strong} at the expense of added optical losses that might limit the achievable finesse \cite{Stambaugh2015From}. 

\section{Acknowledgments}

We thank Aashish Clerk, Peter Gr\"{u}tter, Christian Degen, and Gary Steele for helpful discussions, Bogdan Piciu and Abeer Barasheed for simulation help, Vikramaditya Mathkar and Chris McNally for help with the fiber interferometer, Mattieu Nannini, Don Berry, Jun Li, Lino Eugene, Simon Bernard, and Scott Hoch for fabrication help. T.M. acknowledges support by a Swiss National Foundation Early Postdoc Mobility Fellowship. The authors also gratefully acknowledge financial support from NSERC, FRQNT, the Alfred P. Sloan Foundation, CFI, INTRIQ, RQMP, CMC Microsystems, and the Centre for the Physics of Materials at McGill.

\section*{Appendix I: Fabrication Details}

Fabrication begins by lithographically defining a 1.5-$\um$-thick photoresist mask in the shape of a trampoline on the top surface and transferring it to the nitride with a CF$_4$/CHF$_3$ reactive ion etch (RIE, etch time \SI{45}{s}, RF power \SI{500}{W}\footnote{We now recommend lower power to avoid burning the resist.}, chamber pressure \SI{30}{mTorr}). The remaining resist is left as a protective layer while an array of square openings is patterned into the backside nitride using the same technique. The wafer is then diced into chips of $\SI{15}{mm}\times\SI{15}{mm}$ for handling, each hosting 8 identical devices and one unpatterned ``reference'' membrane (see Fig.~\ref{fig1}(b); reference membrane can be fully etched if desired), and mounted in a chemically-resistant Polytetrafluoroethylene (PTFE) carrier. This carrier\footnote{Designs available upon request.} holds the chips rigidly in a vertical orientation, while allowing liquid to slowly enter and drain via a hole in the bottom; we find it plays a crucial role in device survival during wet chemical processing. The photoresist is stripped in acetone and the newly-exposed silicon's native oxide is removed with a 1-minute 10:1 Hydroflouric (HF) acid dip at room temperature. To release the trampolines, the chips are briefly rinsed in DI water and then transferred to a 45\% potassium hydroxide (KOH) solution at $\SI{60}{^\circ C}$, where the silicon is etched at a rate of 18 $\um$/hr for . This removes the requisite $340$ $\um$ from both sides of the wafer, resulting in the profile of Fig.~\ref{fig1}(a). Faster etches could be achieved at higher temperatures (e.g. $\sim$\SI{30}{\um}/hr at $\SI{75}{^\circ C}$), but we find this significantly reduces device yield, likely due to increased H$_2$ bubble formation \cite{Zubel2004Etch}. We suspect the rising bubbles break the tethers by either directly exposing them to surface tension forces and pressure variations, or by violently shaking the chip (if loosely mounted), thereby dragging the pad (a.k.a.~``the giant sail'') through the solution. While keeping the released devices submerged, the KOH solution is then diluted to 0.1\% of its original strength by iteratively removing existing solution, without exposing the devices to air, and refilling with DI water. This dilution process is repeated with isopropanol to further clean and reduce surface tension. The chips are then transferred to a 10:1 HF solution for \SI{10}{min}, which gently etches $\sim 10$ nm of nitride (from all exposed surfaces) along with any lingering residues \cite{Grutter2015Si3N4}. Finally the chips are transferred to DI water and then methanol for a final rinse before removing and drying on a hotplate at $\SI{85}{^\circ C}$. With this protocol, 6 of the 8 devices in Fig.~\ref{fig1}(b) survived, consistent with a survival rate of $\sim 50$\% for all device types discussed here.

\bibliography{Alles}  

\end{document}